# Nanocrystalline Al-Mg with extreme strength due to grain boundary doping

Simon C. Pun[a], Wenbo Wang[c], Amirhossein Khalajhedayati[b], Jennifer D. Schuler[b], Jason R. Trelewicz[c], Timothy J. Rupert[a,b,*]

[a] Department of Mechanical and Aerospace Engineering, University of California, Irvine, CA 92697, USA
[b] Department of Chemical Engineering and Materials Science, University of California, Irvine, CA 92697, USA
[c] Department of Materials Science and Engineering, Stony Brook University, Stony Brook, NY 11794, USA
[*] Corresponding Author. Tel.: 949-824-4937; e-mail: trupert@uci.edu

Nanocrystalline Al-Mg alloys are used to isolate the effect of grain boundary doping on the strength of nanostructured metals. Mg is added during mechanical milling, followed by low homologous temperature annealing treatments to induce segregation without grain growth. Nanocrystalline Al -7 at.% Mg that is annealed for 1 h at 200 °C is the strongest alloy fabricated, with a hardness of 4.56 GPa or approximately three times that of pure nanocrystalline Al. Micropillar compression experiments indicate a yield strength of 865 MPa and a specific strength of 329 kN·m/kg, making this one of the strongest lightweight metals reported to date.

**Introduction**

Nanocrystalline metals exhibit beneficial mechanical properties such as high strength, prolonged fatigue life, and improved wear resistance [1, 2, 3]. Reducing a material's grain size results in a large increase in the number of grain boundaries, which act as obstacles to dislocation motion [4]. For grain sizes over the range of ~15-100 nm, plasticity is dominated by dislocations that are nucleated from grain boundary sources, travel across the grain while being momentarily pinned at the ends by boundary sites, and finally are absorbed into the opposite grain boundary [5]. While plasticity is still based on dislocation motion, the grain boundary is now heavily involved in the entire process and local grain boundary state should therefore be important. For example, grain boundary relaxation is a process where energy stored during processing in the form of excess grain boundary defects or disorder is released as the boundary transforms towards an equilibrium configuration [6, 7]. This boundary relaxation serves to strengthen a nanocrystalline material, with a study of nanocrystalline Ni-W showing hardness increases of up to 35% compared to the as-deposited material upon low homologous temperature annealing [8]. Molecular dynamics simulations have confirmed such relaxation, showing that relaxed boundaries better resist grain boundary sliding and make dislocation nucleation and propagation more difficult [9, 10].

Grain boundary structure can be intimately tied to grain boundary chemistry as well. Nanocrystalline alloys with segregating dopants have been developed for improved thermal stability [11] and also for the production of materials with controllable grain sizes [12], with predominant theories suggesting that the dopants reduce grain boundary energy [13]. The addition of solutes to interfaces and the reduction of grain boundary energy should also influence the mechanical behavior of nanostructured metals. Extremely high strengths near the theoretical limit were predicted by Vo et al. [14] through molecular dynamics simulations and attributed to heavy doping of the grain boundaries. Ozerinc et al. confirmed that boundary doping can significantly increase the strength of nanocrystalline metals using nanoindentation experiments on Cu, Cu-Nb, and Cu-Fe films [15]. However, grain size varied between samples and solid solution strengthening, while ruled out as the dominant effect, was not explicitly treated and subtracted in Ref. [15].

In this paper, the effect of grain boundary doping on hardness and strength is isolated using nanocrystalline Al and Al-Mg alloys with a constant grain size. By keeping grain size constant, the strengthening contribution from grain size reduction is constant among all testing samples. Low homologous temperature heat treatments are employed to tailor segregation state, which is quantified through X-ray diffraction and contrasted with stable nanocrystalline grain structures created by lattice Monte Carlo (LMC) simulations. The distribution of solute in the grain interior and grain boundary regions is extracted and then used to measure the respective contributions of solid solution strengthening and grain boundary segregation strengthening. We find that boundary segregation has a much larger effect on strength than solid solution addition, which enables the production of alloys with roughly three times the strength of pure nanocrystalline Al. Microcompression experiments were used to confirm the extreme strength of the nanocrystalline Al-Mg alloys, and the results position these materials among the strongest Al alloys to date with a specific strength of 329 kN·m/kg.

**Methods**

Pure nanocrystalline Al and Al-Mg alloys with Mg concentrations up to 7 at.% were produced using a high-energy SPEX 8000M ball mill under Ar atmosphere at 2 x $10^{-3}$ torr. This dopant choice was motivated by the work of Murdoch and Schuh [16], which suggested that Mg has a positive enthalpy of segregation in Al and would therefore have an energetic preference to segregate to the grain boundaries. Selected samples were annealed at 150 °C and 200 °C under vacuum for 1 h, to facilitate solute redistribution toward thermodynamically favored configurations. Structural characterization of both as-milled and annealed specimens was performed by using X-ray diffraction (XRD) in a Rigaku Ultima III Diffractometer with a Cu Kα radiation source operated at 40 kV and 30 mA. XRD profiles were used to verify that all alloys were polycrystalline, face-centered cubic (fcc) solid solutions. Grain size was measured by applying the Scherrer equation to each peak, with the reported values representing the average from all reflections [17]. Since coherent X-ray scattering only occurs from the grain interiors, peak positions were used to quantify the distribution of dopant atoms between grain interior and grain boundary sites, which will be described in detail later. Energy dispersive X-ray spectroscopy (EDS) was performed on a Philips XL-30 field emission scanning electron microscope (SEM) at 10 kV to quantify global alloy composition. Transmission electron microscopy (TEM) samples were made from the powders with the focused ion beam (FIB) lift-out technique, using an FEI Quanta 3D microscope and $Ga^+$ ions. In order to reduce ion beam damage, TEM samples received a final polish with a low power 5 kV beam. Bright field TEM images, selected area electron diffraction (SAED) patterns, EDS profiles, and scanning TEM images were collected on an FEI Titan operating at 300 kV. Analysis of SAED patterns was performed with the Crystallographic Tool Box (CrysTBox) [18].

Hardness was measured using an Agilent G200 nanoindenter with a diamond Berkovich tip, which was calibrated with a standard fused silica specimen. All quoted hardness values are based on a minimum of 30 measurements, determined at an indentation depth of 300 nm and using a constant indentation strain rate of $0.05\ s^{-1}$. Micropillars with average diameters of 6.3 μm and average heights of 16.4 μm were fabricated with automated lathe milling using a focus ion beam (FIB) microscope, following the method of Uchic and Dimiduk [19]. The pillar aspect ratio (height/diameter) of ~2.6 was chosen to follow microcompression testing best practices developed by Zhang et al. [20] to avoid plastic buckling. Yield strength was obtained by performing microcompression on three individual micropillars with a flat punch tip in the same nanoindenter and then using the 0.7% yield strain offset criterion of Brandstetter et al. [21].

Solute distributions and grain boundary formation energies for stable nanocrystalline states in Al-Mg alloys were explored as a function of alloy composition and grain size using the LMC formalism. The LMC methodology elucidates the lowest free energy state of an alloy from a configurational space that incorporates chemical mixing (i.e. as originally captured by the Ising model [22]) collectively with grain boundaries. The latter was enabled by assigning each lattice site a grain allegiance in addition to a chemical identity with nearest-neighbor bond energies parameterized via the Regular Nanocrystalline Solution (RNS) model [23], which prescribes unique energies to crystalline (grain interior) and grain boundary bonds. Pioneering work was conducted on tungsten alloys that identified nanocrystalline stable states in refractory metal alloys [24], and the technique was recently applied to compare predicted stable and unstable configurations in different binary iron alloys [25].

Seven alloy compositions were investigated including 1, 2, 5, 7, 10, 15, and 20 at.% Mg, and each simulation cell was constructed to contain 400 x 400 x 6 sites on an fcc lattice. The chemical identity and grain allegiance of each site was randomly assigned by initializing the simulation at 10,000 K, followed by cooling to the equilibration temperature ($T_{eq}$) at a rate of $-\frac{(T_{step}-T_{eq})}{1000}$. A single equilibration temperature of 300 K was employed and each configuration was evolved at this temperature using the Metropolis method [26]. The lowest free energy state was identified using a convergence rate defined by $\bar{S} = \frac{\Delta E}{\Delta l}$ where $\Delta E$ represents the internal energy difference over a prescribed number of Monte Carlo steps defined by $\Delta l$; herein, a constant value of 2000 was assigned for $\Delta l$. The system was evolved to a state where the convergence rate was $< 10^{-6}$, and followed by an additional 50,000 Monte Carlo steps to confirm the lowest free energy state was in fact achieved.

The internal energy of the system for a given configurational state was determined by summing over all bond energies for nearest-neighbor bonds following the implementation of the RNS model for the LMC framework [27]. This energy parameterization distinguished like and unlike bonds in both the crystalline lattice and grain boundary regions via distinct interaction energies denoted $\omega_c$ and $\omega_{gb}$, respectively. The bulk interaction energy of $\omega_c = 0.13$ kJ/mol was ascertained from the dilute heat of mixing, $\Delta H_{mix}$, for Al-Mg of 1.607 kJ/mol [28], which enabled the energy of the unlike bonds in the crystalline lattice ($E^c_{Al-Mg}$) to be determined using like bond energies of $E^c_{Al-Al}= 135$ kJ/mol and $E^c_{Mg-Mg}= 9$ kJ/mol [29]. Given these like bond energies, a dilute heat of segregation, $\Delta H_{seg}$, of 12.744 kJ/mol [28], and grain boundary energy penalty, $\frac{2\Omega \gamma_o}{zt}$, for the solvent and solute of 2.14 and 2.8 kJ/mol, respectively, the grain boundary interaction parameter was determined to be $\omega_{gb} = -1.73$ kJ/mol. The grain boundary bond energies were computed using the above values, thus providing the requisite energy parameterization for calculating the total internal energy of each configurational state.

**Results and Discussion**

Hardness is shown as a function of global alloy composition in Figure 1 for nanocrystalline Al-Mg alloys with a mutually consistent grain size of 24 nm. The sample containing 7 at.% Mg exhibited a hardness of 4.19 GPa whereas the hardness of nominally pure nanocrystalline Al was 1.61 GPa, which demonstrates the dramatic increase in hardness achieved through alloying. Annealing at low homologous temperatures further increased the hardness of the alloys, with a maximum hardness of 4.56 GPa measured following the 200 °C heat treatment. This particular alloy is roughly three times as strong as pure nanocrystalline Al and is among the strongest Al-based materials that can be found in the literature, surpassing the hardness of commercial Al alloys [30]. With the XRD measurements confirming a consistent average grain size of 24 nm for all samples, strengthening from grain size effects were ruled out of the analysis.

Additional characterization was performed to ensure that there were no second phase precipitates in the microstructure. Figure 2(a) presents XRD profiles from the pure Al sample and the Al-7 at.% Mg sample that was annealed at 200 °C, showing that only peaks consistent with an fcc phase are observed. However, XRD probes a relatively large volume of material and cannot be used to rule out nanoscale precipitates, meaning TEM characterization was also required. Figure 2(b) and (c) present a bright field TEM image and a SAED pattern taken from the Al-7 at.% Mg sample that was annealed at 200 °C, respectively. A nanocrystalline grain structure with an average grain size of ~26 nm is observed, consistent with the measurements taken from XRD. Figure 2(c) shows that only diffraction rings from an fcc Al-rich phase are observed, with the red curve in the figure showing the radially averaged intensity. On this same figure, we purposely plot the location of Mg diffraction rings to show that there is no precipitation of Mg-rich hexagonal close packed (hcp) particles. Darling et al. [31] observed that dopant-rich second phase particles can precipitate at grain boundaries in Cu-Ta after annealing treatments, but we do not observe such a phenomenon here. As a whole, our characterization results show that we only have a single fcc phase in our nanostructured material, so the Mg atoms must be either incorporated into the lattice as solutes or segregated to the boundaries. A scanning TEM image from the same Al-7 at.% Mg sample annealed at 200 °C is presented in Figure 3(a), with a representative EDS line scan shown in Figure 3(b). The composition measurements fluctuate around the average value of 7 at.% Mg, with low compositions associated with the depleted crystal interiors and higher compositions associated with enriched grain boundaries.

To develop a complete picture of solute strengthening, the relative contributions of grain size, solid solution strengthening, and grain boundary segregation were extracted. The total strength of a nanocrystalline alloy will be comprised of the strength of the pure material plus a contribution from the dopant addition:

$$H_{Alloy} = H_{Pure} + \Delta H_{Doping} \tag{1}$$

where $H_{Alloy}$ is the measured alloy hardness, $H_{Pure}$ is the measured baseline nanocrystalline Al hardness, and $\Delta H_{Doping}$ is the solute doping contribution. As discussed previously, Mg dopants can be incorporated either into the lattice as a solid solution or into the grain boundaries, with the degree of segregation depending on the relative enthalpies of mixing and segregation. In both cases, this may have a strengthening effect on the alloy. As such, the strengthening effect of chemical dopants can be further decomposed into solid solution and boundary doping contributions:

$$\Delta H_{Doping} = \Delta H_{SSS} + \Delta H_{GB\ Seg.} \tag{2}$$

where $\Delta H_{SSS}$ is the change in hardness from solid solution strengthening and $\Delta H_{GB\ Seg.}$ is the change in hardness from grain boundary doping. The solid solution strengthening term was treated using the formulation introduced by Rupert et al. [32], which added the contribution of dislocation pinning in nanoscale grains to the traditional solid solution formulations such as the Fleischer formulation [33] that have been used extensively to describe coarse-grained metals. Calculation of this solid solution strengthening only requires knowledge of the elastic modulus and Burgers vector of each base element, coarse-grained solid solution strengthening data, and information about the particular nanocrystalline alloy of interest such as grain size and grain interior composition. Changes in yield strength from solid solution strengthening theory can be converted to hardness values using a modified Tabor relation ($H = 3.8\sigma_y$, where H is hardness and $\sigma_y$ is yield strength) introduced by Dalla Torre et al. [34]. After subtracting the strengthening contributions from grain size and solid solution doping, the remaining strength can be attributed to grain boundary segregation effects.

In order to separate the two doping effects, the concentration of solutes at the grain interior as well as grain boundary were calculated from the global composition. If Mg atoms are incorporated as substitutional solutes, the Al lattice will swell and the lattice constant measured by XRD will increase. On the other hand, Mg atoms that go to grain boundary sites will not affect the peak positions measured by XRD. Since the global composition is known and grain size is constant, the lattice constant measurements can be used to distinguish the grain boundary and grain interior (lattice) compositions. The lattice parameter measurements are shown in Figure 4 for both the as-milled and heat-treated alloys as a function of global Mg concentration. The alloys follow the expected linear trend of Vegard's law [35] up to a concentration of 2 at.% Mg, indicating that the Mg was incorporated as solid solution defects at low concentrations. Further addition of Mg increases the lattice parameter, but at a reduced rate relative to the predictions for a perfect solid solution. This suggests that some of the additional Mg was incorporated into the lattice but there was a slight preference for segregation even in the as-milled state. Annealing promoted further solute enrichment of the grain boundaries as captured by the decreasing lattice parameter.

The distribution of Mg in the alloy can be described in terms of the global composition ($C_{global}$), which represents a weighted sum of the grain boundary and lattice concentrations:

$$C_{Global} = V_{GB}\, C_{GB} + V_{Lattice}\, C_{Lattice} \qquad (3)$$

where $V_{GB}$ and $C_{GB}$ are the volume fraction and composition of grain boundaries, respectively, and $V_{Lattice}$ and $C_{Lattice}$ are the volume fraction and composition of lattice, respectively. Palumbo et al. [36] introduced a geometric model that modeled grains as fourteen-sided tetrakaidecahedron in order to estimate the volume fraction of material located at intercrystalline regions. For a 24 nm grain size and a 1 nm grain boundary width, the grain boundary and lattice volume fractions would be 0.12 and 0.88, respectively. $C_{Lattice}$ can be obtained from the measured lattice constant while $C_{Global}$ is given by the initial powder mixture and confirmed by EDS. This leaves only one unknown, $C_{GB}$, which was then calculated from Eqn. 3. Using this method, the grain interior and grain boundary concentrations were extracted, with a few representative examples shown as inset schematics in Figure 4.

The change in hardness from solid solution strengthening is presented in Figure 5(a) as a function of the lattice concentration of Mg, calculated based on the work of Rupert et al. [32] and using coarse-grained data taken from Lee et al. [37]. While hardening occurred with increasing lattice concentration, the contribution to the overall hardness was relatively small. A maximum hardness increment of 0.38 GPa was found when 4.2 at.% Mg was mixed into the Al lattice, which amounted to only ~9% of the total hardness. The hardness contribution from grain boundary doping was then obtained by subtracting out hardness increments from grain size and solid solution effects following Eqns. 1 and 2, with the results presented in Figure 5(b) as a function of grain boundary Mg concentration. In this case, there was a maximum hardness increment of 2.58 GPa when 31 at.% Mg is mixed into the grain boundaries, which amounted to ~57% of the total hardness. Consequently, not only did the grain boundary doping effect dominate over solid solution strengthening in the nanocrystalline Al-Mg alloys, but it also surpassed the strengthening increments resulting from more than the grain size effect for these alloys. Our observations are in agreement with the work of Ozerinc et al. [15] on nanocrystalline Cu-Nb alloys, who also found that grain boundary doping overshadowed grain size strengthening. The trend in Figure 5(b) rapidly increases as grain boundaries initially became enriched in Mg; however, the subsequent plateau suggests that grain boundary segregation strengthening saturated despite the continued increase in $C_{GB}$.

We can hypothesize that this strengthening trend is related to grain boundary energy, but such a variable is not easily tractable experimentally. Therefore, the mechanisms responsible for the plateau in grain boundary segregation strengthening were investigated using LMC simulations in the Al-Mg system for global compositions spanning the range explored with the experimental alloys. Stable nanocrystalline states were achieved for all compositions over the range of 1–20 at.% Mg and the equilibrated structures are illustrated in Figure 6. In this figure, the color employed in the upper panels delineate grains while Mg dopants are represented by the black dots; color is eliminated in the lower figure panels for a clearer presentation of the segregation state. At low solute contents, i.e. $C_{Global} \leq 2$ at.% Mg, solute atoms predominantly segregated to the grain boundaries with only a few Mg atoms apparent within the grains. This finding is consistent with the observation that annealing of the lower concentration alloys promotes segregation from the metastable solid solution achieved through ball milling alone, which produces alloys that are far from equilibrium due to the high milling energy. Increasing the solute content enhanced the grain interior solute concentration while no discernible changes were qualitatively apparent in the grain maps. Despite this implied change in the solute distribution, the grain size determined from the total grain boundary area under a spherical grain assumption exhibited only a minimal decrease from 24 to 20 nm over the entire composition range of 1–20 at.% Mg.

The segregation isotherm shown in Figure 7(a) was constructed to quantify the solute distribution in the equilibrium computational structures, as this can be compared directly with the experimental lattice parameter analysis. A sharp increase in the grain boundary solute content occured over the global composition range of 1–5 at.% Mg, confirming that Mg atoms predominately segregate to grain boundaries at low solute contents. The increase to 7 at.% Mg was attended by a maximum in the grain boundary solute content of approximately 32 at.% Mg, which is comparable to the grain boundary composition determined from lattice parameter analysis for the experimental sample with the highest concentration. Further increasing the global composition toward 20 at.% produced a decrease in the grain boundary solute content relative to the grain interior, which is best characterized using the Gibbsian solute excess [38], $\Gamma$, defined as:

$$\Gamma = \frac{1}{A_{gb}} \left( N_{gb}^{Mg} - N_c^{Mg} \left( \frac{N_{gb}^{Al}}{N_c^{Al}} \right) \right) \quad (4)$$

where $A_{gb}$ represents the grain boundary area and $N$ the number of Mg or Al atoms in the grain boundary or grain interior denoted by the subscripts 'gb' and 'c', respectively. The intrinsic dependence of $\Gamma$ on grain boundary area accounts for variations in grain size, which are not explicitly captured in the segregation isotherm shown in Figure 7(a). The grain boundary solute excess in the equilibrated structures is shown in Figure 7(b) over a truncated range of global solute contents that coincides with the experimental alloy compositions.

A sharp increase in $\Gamma$ over the global composition range of 1–5 at.% Mg aligns with the strengthening regime in Figure 1, and the subsequent maximum at 7 at.% Mg also coincides with the beginning of a hardness plateau in Figure 5. However, analysis of the solute distribution alone does not capture the implications of the various segregation states on the nature of the grain boundaries, namely how doping alters grain boundary energy. Here, we employ the grain boundary formation energy for a closed system following Chookajorn and Schuh [27]:

$$\gamma = \frac{\Delta E_{defect} - \Delta E_{ideal}}{A_{gb}} \quad (5)$$

where the numerator describes the difference in the formation energy of a polycrystal and single crystal with identical chemical order. Negative values of γ in Figure 7(b) are a consequence of the grain boundary segregation state achieved in each equilibrium structure. The reduction in γ with increasing global composition signaled a transition to a more stable grain boundary configuration, which has previously been connected to strengthening effects in nanocrystalline alloys via grain boundary relaxation [14]. However, this effect was exhausted at a global composition of approximately 5 at.% Mg, manifesting as plateaus in both the γ and *ΔH*<sub>GB Segreg.</sub> trends. The strengthening increments due to grain boundary segregation can thus be understood in the context of solute enrichment augmenting the nature of the grain boundaries as captured by the reduction in the grain boundary formation energy with increasing global composition.

With a more complete understanding of grain boundary segregation strengthening, attention was turned to uniaxial deformation to confirm the yield strength of these alloys. An SEM image of a taper-free micropillar fabricated from the hardest alloy is shown in Figure 8(a), with the corresponding stress-strain curves from compression of three identical pillars shown in Figure 8(b). The pillar experiments demonstrated that this alloy exhibits a yield strength of 865 ± 39 MPa, which is significantly greater than the upper limit of ~700 MPa reported for traditional age-hardened Al alloys [39] and for most nanostructured or ultra-fine grained Al alloys created by severe plastic deformation [40, 41]. The few examples of nanostructured Al alloys that are stronger typically rely on precipitation hardening for the added strength. For example, the dispersion of fine $Al_7Cr$ and Cr particles in an Al-20 wt.% Cr alloy produced a yield strength of 1104 MPa [42]. Only the hierarchically nanostructured Al 7075 alloy reported by Liddicoat et al. [43] maintains a single phase solid solution and reports a yield strength marginally greater (978

MPa) than the alloys studied here. Interestingly, these authors also reported a strengthening effect that was dependent on grain boundary doping, showing that nanometer-scale intergranular solute structures were prevalent in their material. In addition to being among the strongest Al alloys reported, the materials presented here take advantage of a very light dopant to achieve the enhanced strength, with Mg being only ~64% as dense as Al. This translates to an extremely high specific strength of 329 kN·m/kg for the nanostructured Al-7 at.% Mg alloy that was annealed at 200 °C for 1 h. Many high-strength Al alloys employ heavier dopants that rapidly reduce the specific strength. For example, the nanostructured Al-20 wt.% Cr with the highest reported strength in the literature [42] only exhibits a specific strength of 275 kN·m/kg. Since Al is widely selected for its high strength-to-weight ratio, dopant selection is critical in the design of high-strength lightweight metal alloys.

**Conclusions**

In conclusion, the strengthening effect from grain boundary doping has been isolated and systematically studied in Al-Mg nanocrystalline alloys. Segregating grain boundary dopants can significantly increase hardness and yield strength by augmenting the nature of the grain boundaries as substantiated by the reduction in the grain boundary formation energy. Mg atoms were found to segregate to grain boundary sites due to the high grain boundary segregation enthalpy of this system, with segregation further enhanced when the materials were annealed. Our nanocrystalline Al-7 at.% Mg alloy with a 24 nm grain size exhibited a hardness of 4.56 GPa, a yield strength of 865 MPa, and a specific strength of 329 kN·m/kg. As a whole, these results provide evidence that grain boundary segregation is a promising route to the creation of alloys with extreme strength.

**Acknowledgements**

This study was supported by the U.S. Army Research Office under Grant W911NF-16-1-0369. J.R.T and W.W. acknowledge support from the National Science Foundation through Grant DMR-1410941.

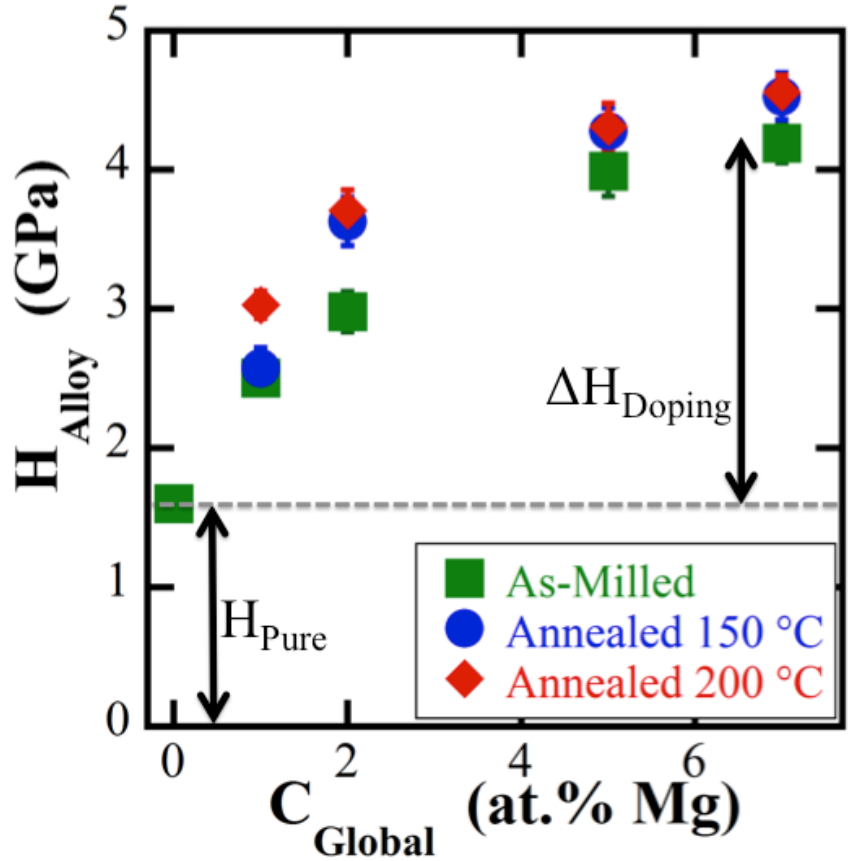

Fig. 1. Hardness values of Al-Mg alloys plotted against the global Mg concentration. While the as-milled alloys show high hardness, the annealing provides relaxation that further hardens the alloys. It is also obvious that strengthening from grain size reduction is much less than that from doping.

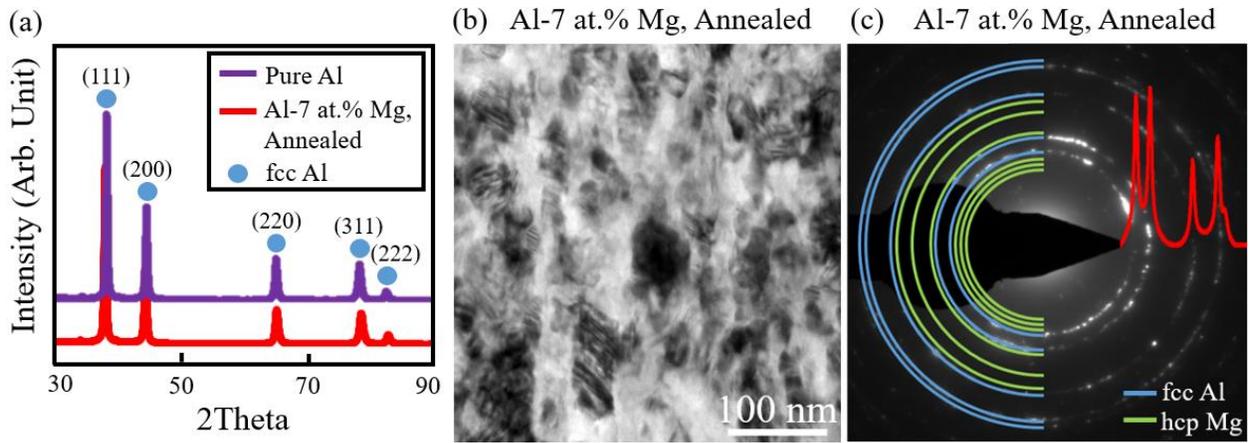

Fig. 2. (a) XRD line profiles for pure nanocrystalline Al and Al-7 at.% Mg annealed at 200 °C. (a) Bright field TEM image and (c) TEM diffraction pattern showing a polycrystalline grain structure with only an fcc phase present for the Al-7 at.% Mg sample annealed at 200 °C.

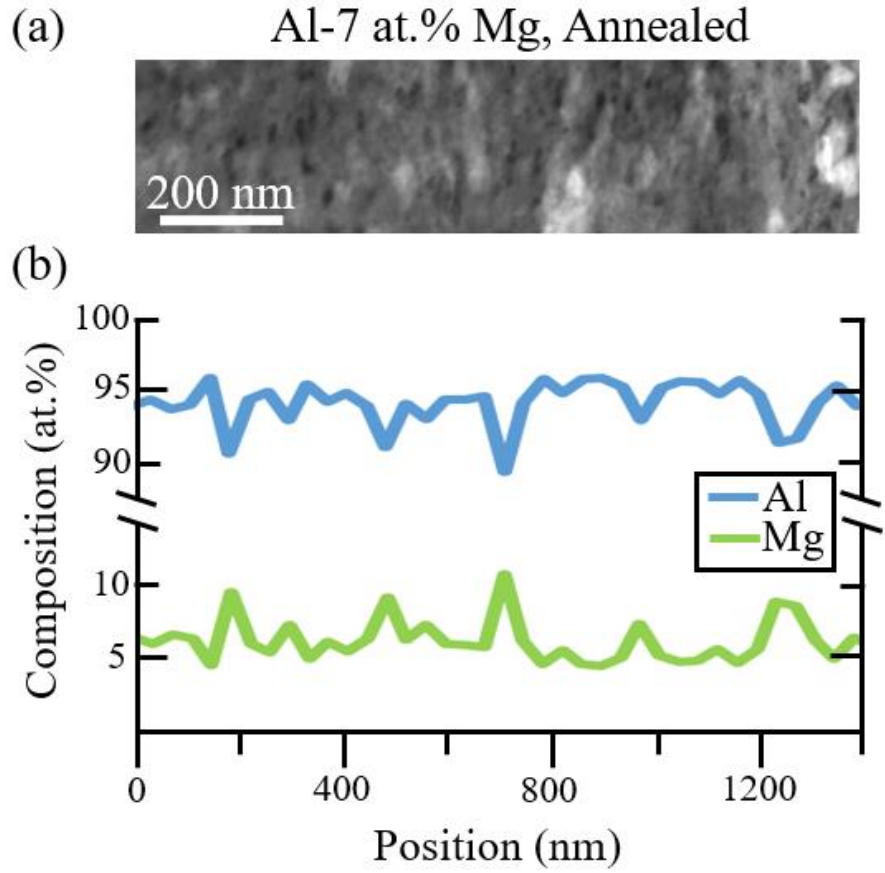

Fig. 3. (a) Scanning TEM image and (b) EDS line profile scans of the Al-7 at.% Mg sample annealed at 200 °C.

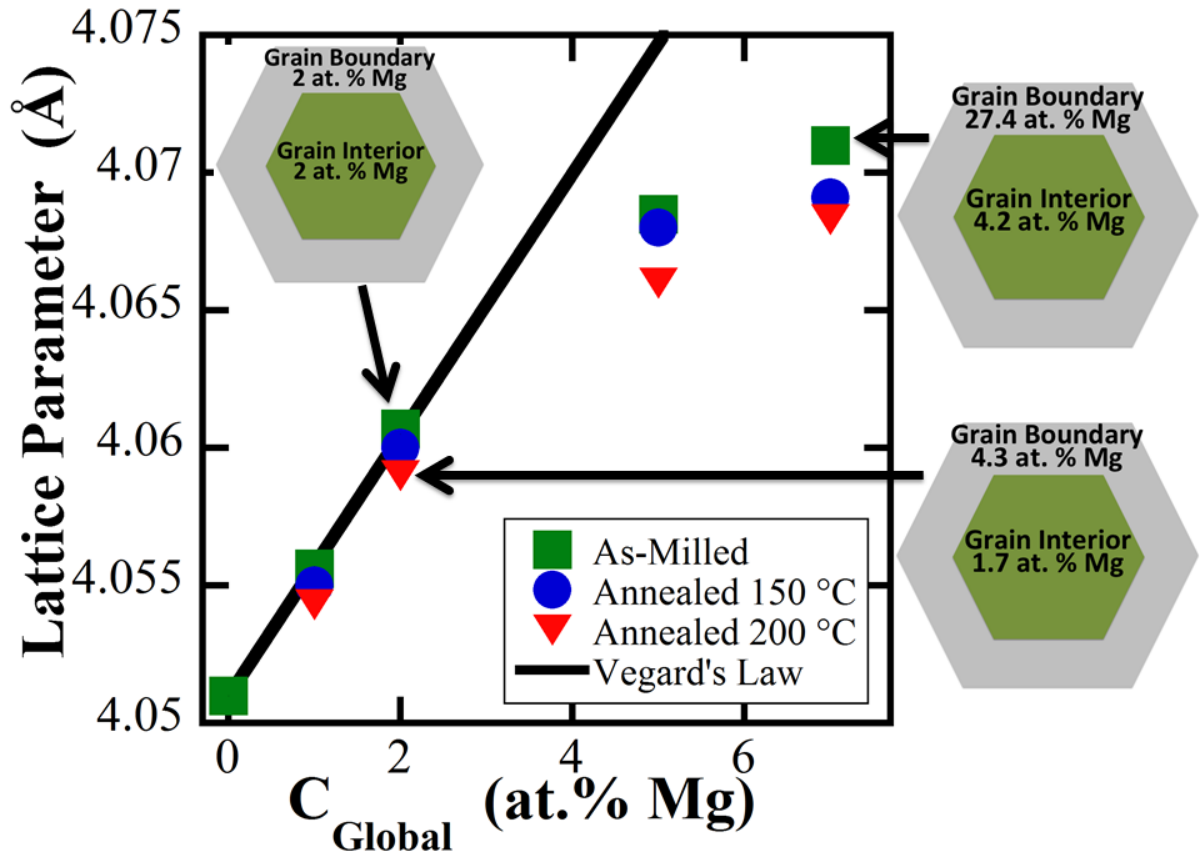

Fig. 4. Lattice parameter is shown as a function of global Mg concentration. The experimental data deviates from theoretical Vegard's Law at solute concentration above 2 at. %, indicating an oversaturation of solute in the lattice. Mg atoms are observed to segregate out of the lattice into the grain boundary as a result of annealing. The schematics of chemical concentration distribution show the grain interior region in green and boundary region in grey.

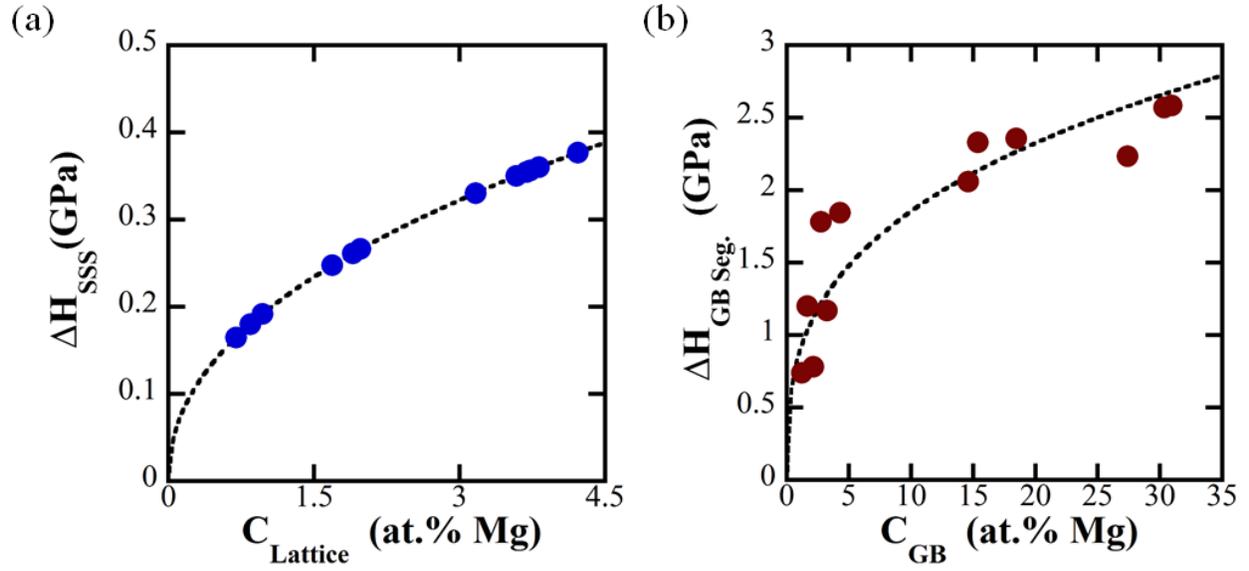

Fig. 5. Hardness increases from various strengthening mechanisms plotted against Mg concentration in (a) the lattice and (b) the grain boundary. While there are strengthening effects from solid solution and grain size reduction, the increase in strength due to grain boundary doping makes the largest contribution in heavily doped Al-Mg nanostructured alloys.

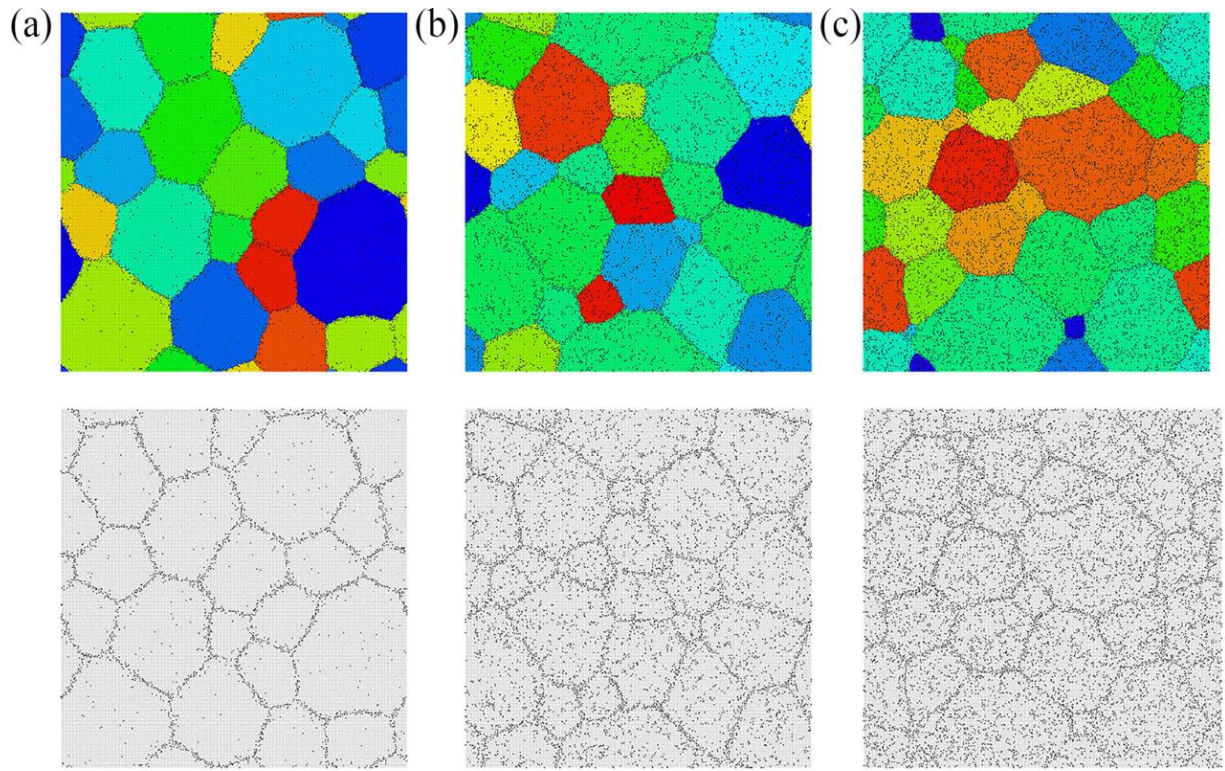

Fig. 6. Grain allegiance and solute distribution maps for Al-Mg alloys containing (a) 2 at.%, (b) 5 at.%, and (c) 7 at.% Mg with nominally equivalent grain sizes. The majority of the Mg atoms initially segregate to the grain boundaries at low solute contents, and subsequent Mg additions are accommodated in the grain interiors due to solute saturation of the grain boundaries.

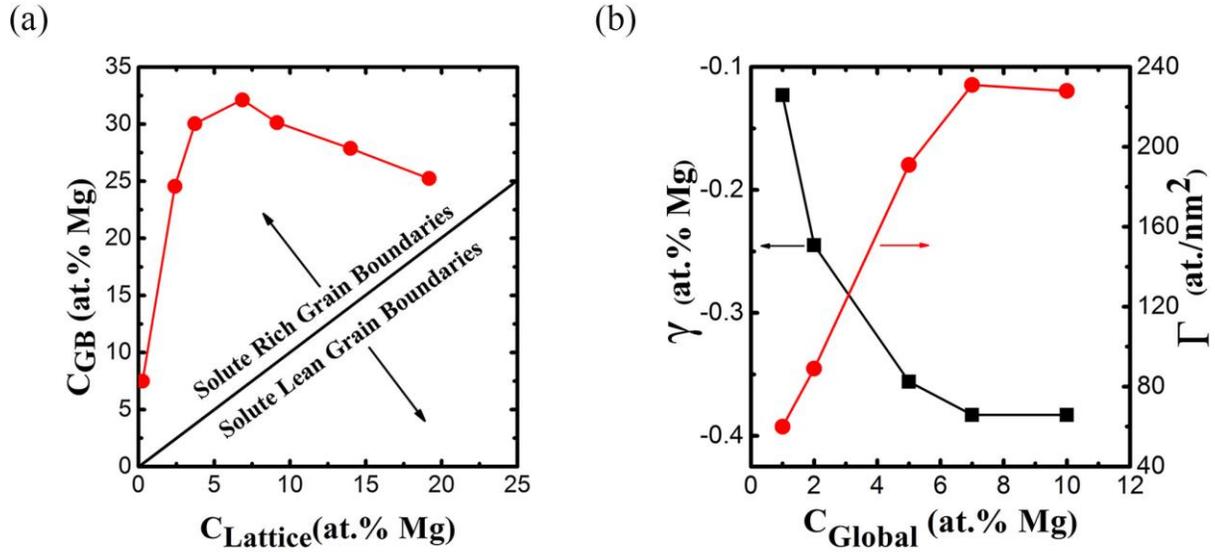

Fig. 7. Quantitative results for the Al-Mg alloys from the LMC simulations including (a) segregation isotherm and (b) corresponding solute excess and grain boundary formation energies for the first five global solute contents from (a) that span the experimental alloy compositions. The minimum in the grain boundary formation energy derives from solute saturation of the grain boundaries and coincides with the plateau in grain boundary segregation strengthening.

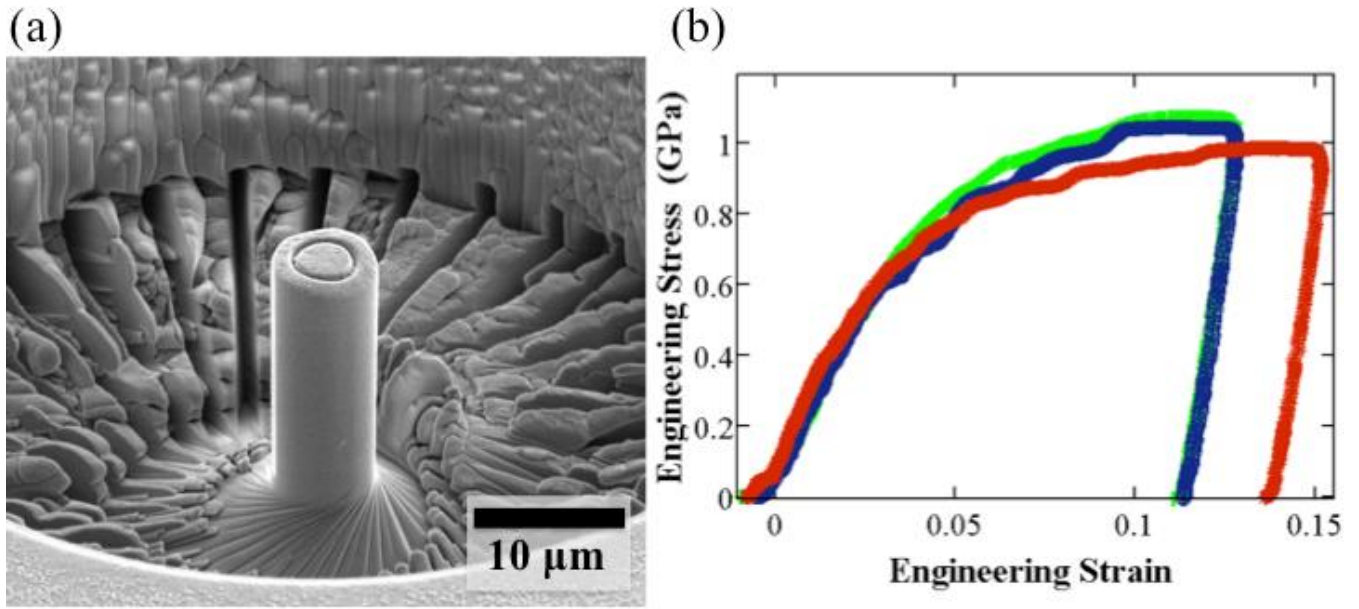

Fig. 8. (a) SEM image of a representative micropillar in Al–7 at. % Mg, that is subsequently used for compression testing. (b) Stress-strain curves obtained from microcompression testing of three individual micropillars.